%% file: 00_main.tex
\documentclass[conference]{IEEEtran}
\IEEEoverridecommandlockouts

\input{99-preamble}

\newboolean{showcomments}
\setboolean{showcomments}{false}

\ifthenelse{\boolean{showcomments}}
{\newcommand{\nbc}[3]{
 {\colorbox{#3}{\bfseries\sffamily\scriptsize\textcolor{white}{#1}}}
 {\textcolor{#3}{\sf\small$\blacktriangleright$\textit{#2}$\blacktriangleleft$}}
 }
}
{\newcommand{\nbc}[3]{}
 }
\newcommand\lili[1]{\nbc{Lili}{#1}{purple}}
\newcommand\sogol[1]{\nbc{Sogol}{#1}{teal}}

\begin{document}

\title{\classifier: A Framework for Producing Classifiers of Security-Related Issue Reports\\
}

\author{
\IEEEauthorblockN{Sogol Masoumzadeh}
\IEEEauthorblockA{\textit{Electrical \& Computer Engineering} \\
\textit{McGill University}\\
Montréal, Canada \\
Sogol.masoumzadeh@mail.mcgill.ca}
\and
\IEEEauthorblockN{Yufei Li}
\IEEEauthorblockA{\textit{Electrical \& Computer Engineering} \\
\textit{McGill University}\\
Montréal, Canada \\
Yufei.li2@mail.mcgill.ca}
\and
\IEEEauthorblockN{Shane McIntosh}
\IEEEauthorblockA{\textit{Cheriton School of Computer Science} \\
\textit{University of Waterloo}\\
Waterloo, Canada \\
Shane.mcintosh@uwaterloo.ca}
\and
\IEEEauthorblockN{\hspace{2cm}Dániel Varró}
\IEEEauthorblockA{\hspace{2cm}\textit{Computer \& Information Science} \\
\hspace{2cm}\textit{Linköping University}\\
\hspace{2cm}Linköping, Sweden\\
\hspace{2cm}Daniel.varro@liu.se}
\and
\IEEEauthorblockN{\hspace{2cm}Lili Wei}
\IEEEauthorblockA{\hspace{2cm}\textit{Electrical \& Computer Engineering} \\
\hspace{2cm}\textit{McGill University}\\
\hspace{2cm}Montréal, Canada\\
\hspace{2cm}Lili.wei@mcgill.ca}
}

% \IEEEpubid{The manuscript has been accepted for publication at SANER 2026!}
% \maketitle
\twocolumn[
\begin{@twocolumnfalse}
\begin{center}
\fbox{\parbox{0.95\textwidth}{\centering\small \copyright~2025 IEEE. Author pre-print copy. The manuscript has been accepted for publication at SANER 2026!}}
\end{center}
\vspace{1em}
\maketitle
\end{@twocolumnfalse}
]

\input{01_abstract}
\input{01_introduction}

\input{02_background}
\input{03_study_design}
\input{04_experiment_design}
\input{04_results}
\input{05_implications}
\input{06_threats_to_validity}
\input{07_related_work}
\input{08_conclusion}

\input{09_data_availability}
\balance
\bibliographystyle{IEEEtran}
\bibliography{99_references} 

\end{document}

%% file: 99-preamble.tex
\input{preamble/commands}

\input{preamble/packages}

%% file: preamble/commands.tex
\newcommand{\secissue}{security-related issues\xspace}
\newcommand{\secissuesingle}{security-related issue\xspace}

\newcommand{\dnn}{DNN\xspace}
\newcommand{\dnns}{DNNs\xspace}
\newcommand{\classifier}{\textsc{SeBERTis}\xspace}
\newcommand{\labels}{pseudo-labels\xspace}
\newcommand{\keywords}{Semantic Surrogates\xspace}
\newcommand{\mlm}{MLM\xspace}
\newcommand{\mlms}{MLMs\xspace}
\newcommand{\baselinecount}{five\xspace}

\newcommand{\wild}{in-the-wild\xspace}
\newcommand{\openai}{OpenAI\xspace}
\newcommand{\meta}{Meta}
\newcommand{\gpt}{GPT\xspace}
\newcommand{\llama}{Llama\xspace}
\newcommand{\bert}{BERT\xspace}
\newcommand{\hyper}{hyper-parameter\xspace}
\newcommand{\sota}{state-of-the-art\xspace}
\newcommand{\adapt}{\textsc{AdaptIRC}\xspace}
\newcommand{\fast}{\textsc{fasttext}\xspace}
\newcommand{\random}{random keywords\xspace}

\newcommand\circled[1]{%
  \Circled{#1}}

\usepackage[most]{tcolorbox}
\newtcolorbox{myframe}[1][]{
  enhanced,
  sharp corners, % Ensures no rounded edges
  colback=white,
  boxrule=0.6pt,
  colback=gray!6, % Very light gray background for subtle shading
  #1
}

\newcommand{\reqone}{\textbf{RQ1:} How good is \classifier in training \dnns as \secissuesingle classifiers?\xspace} 
\newcommand{\reqoneone}{\textit{RQ1.1:} What is the best classification performance an \classifier-trained \dnn can achieve?\xspace} % comparison of the 1k dataset and 10k dataset, set epochs=3 for the 
\newcommand{\reqonetwo}{\textit{RQ1.2:} What are the optimal \hyper settings for achieving the best classification performance?\xspace} %epochs and lr

\newcommand{\reqonethree}{\textit{RQ1.3:} How do \classifier-trained \dnns perform compared to \sota issue classifiers?\xspace} % comparison of the 1k dataset and 10k dataset, 
\newcommand{\reqtwo}{\textbf{RQ2:} How important is the choice of masking vocabulary?\xspace} % comparison of the classifier performance when mutual keywords are included and when they are removed. 6 mutual keywords

\newcommand{\reqthree}{\textbf{RQ3:} How well does a \classifier-trained \dnn perform as an off-the-shelf \secissuesingle classifier? \xspace} % thinking of creating a 1k dataset (from the new dataset Mary has gathered, ensure there are not any mutual issues between the 10k dataset and the new 1k one. Pre-process full. Just get the trained classifier and apply it on the dataset to get the accuracy. 

\newcommand{\justrqone}{RQ1\xspace}
\newcommand{\justrqtwo}{RQ2\xspace}
\newcommand{\justrqthree}{RQ3\xspace}

\newcommand{\justrqoneone}{RQ1.1}
\newcommand{\justrqonetwo}{RQ1.2}
\newcommand{\justrqonethree}{RQ1.3}

\newcommand{\imp}{\textit{Imp.}\xspace}

%% file: preamble/packages.tex
\usepackage[utf8]{inputenc}
\usepackage{cite}
\usepackage{amsmath,amssymb,amsfonts}
\usepackage{algorithmic}
\usepackage{graphicx}
\usepackage{subcaption}
\usepackage{textcomp}
\usepackage{xcolor}
\usepackage{xspace}
\usepackage{booktabs}
\usepackage{multirow}
\usepackage{circledsteps}
\usepackage{soul}
\usepackage{balance}
\usepackage[colorlinks=false, linkcolor=blue, citecolor=blue, urlcolor=blue, pdfborder={0 0 1}, citebordercolor=green]{hyperref}

%% file: 01_abstract.tex
\begin{abstract}
Monitoring issue tracker submissions is a crucial software maintenance activity. A key goal is the prioritization of high risk security-related bugs. If such bugs can be recognized early, the risk of propagation to dependent products and endangerment of stakeholder benefits can be mitigated. To assist triage engineers with this task, several automatic detection techniques, from machine learning (ML) models to prompting large language models (LLMs), have been proposed. Although promising to some extent, prior techniques often memorize lexical cues as decision shortcuts, yielding low detection rate specifically for more complex submissions. %with indirect hints to underlying vulnerabilities. 
As such, these classifiers do not yet reach the practical expectations of a 
% In return, their application as 
real-time detector of \secissue. 
% detectors become impractical. 
To address these limitations, %of prior techniques, 
we propose \classifier, a framework to train deep neural networks (\dnns) as classifiers independent of lexical cues, so that they can confidently
% and accurately 
detect fully unseen \secissue. \classifier capitalizes on fine-tuning bidirectional transformer architectures as masked language models (\mlms) on a series of semantically equivalent vocabulary to prediction labels (which we call \keywords) when they have been replaced with a mask. %the \textsf{[MASK]} token. 
Our \classifier-trained classifier achieves a 0.9880 F1-score in detecting \secissue of a curated corpus of 10,000 GitHub issue reports, substantially outperforming \sota issue classifiers, with 14.44\%-96.98\%, 15.40\%-93.07\%, and 14.90\%-94.72\% higher detection precision, recall, and F1-score over ML-based baselines. Our classifier also substantially surpasses LLM-based baselines, with an improvement of 23.20\%-63.71\%, 36.68\%-85.63\%, and 39.49\%-74.53\% for precision, recall, and F1-score, respectively. Finally, our %\classifier{}-trained 
classifier demonstrates a high confidence in detecting recently submitted \secissue, achieving 0.7123, 0.6860, and 0.6760 precision, recall, and F1-score, comparable to those of promoting LLMs,
making it a practical tool for real-time issue report triage. 
% submissions. 
\end{abstract}


%% file: 01_introduction.tex
\section{Introduction}\label{sec:introduction}
% Software maintenance constitute of bug fixing, new features implementation, and alike activities with the goal of persevering the quality of software products after their release~\cite{adams206release}.
\lili{I think the previous sentence is too remote. We could directly start from the motivation of classifying issue reports.}\sogol{Done!}
% Software maintenance is an essential step in the software release pipeline, ensuring the persistence of the software product quality through time. Activities such as the implementation of new features and bug fixing are all a part of software maintenance~\cite{adams206release}. 
Monitoring issue tracker submissions plays a vital role in software maintenance by enabling triage engineers prioritizing bug reports based on their sensitivity, severity, and risk level, thereby ensuring that the most critical defects such as \textit{\secissue} are addressed first by engineers~\cite{yang2014towards}. 
% Among others, bug fixing can be done through the triage of issue reports. Specifically, a newly submitted issue report to a software product's issue tracker is evaluated by a triage engineer, prioritizing sever issues such as security-related defects or \textit{software vulnerabilities}~\cite{yang2014towards}. 
Triageing issue reports for popular software products, with their issue trackers attracting a large volume of submissions on a daily basis, can be time-consuming and laborious~\cite{kallis2021predicting}. At the same time, the difficulty of correctly reporting \secissue grows in folds for more complex submissions that describe unexpected software behavior, a by-product of the underlying latent vulnerability, rather than including explicit definitions or familiar security-related terminology~\cite{sawadogo2021early}. Thus, incorporating automatic detection tools greatly assist triage engineers to accurately and efficiently prioritize \secissue. 
% At the same time, due to limited subject-matter expertise 
\lili{I think it is too ``harsh'' to say that the triage engineers have limited subject-matter expertise. It is unclear what ``subject-matter expertise'' actually refer to. The knowledge of the specific software or security?}
% triage engineers are often challenged to correctly distinguish more complicated \secissue containing descriptions of by-product software behavior 
\lili{Here, what is by-product software behaviour?} 
% of an undiscovered vulnerability~\cite{sawadogo2021early}.  
% Meanwhile, the inherently complicated nature of security-related topics combined with frequent lack of subject matter expertise, makes the identification of \secissue a challenging task for the triage engineer~\cite{sawadogo2021early}. 
% The difficulty of correctly reporting \secissue grows in folds for popular software products, with their issue trackers attracting a large volume of submissions on a daily basis. 
\lili{We may consider to mention this reason first? Also, we may add a reference for this statement.}\sogol{Rewrote the above paragraph to address all your three comments, please give a look.}
%
% Moreover, popular projects attract a large volume of new issue report submissions on a daily basis, making the manual analysis of them even a much more difficult tasks for the triage engineer. Hence, automatic techniques for detecting potentially, sever, \secissue has gained an attraction among software developers.

Earlier attempts for automatic detection of \secissue use their structured information, such as the GitHub assignee field, to mine their textual information and vectorize them based on their vocabulary frequencies~\cite{zhou2016combining, merten2016software}. %Text-based 
\textit{Machine learning (ML)} classifiers are then used to classify the issue reports based on the differences of their numerical vector representations. The accuracy of these text-mining inspired techniques depends on the availability of structured information in the issues. Meanwhile, many of these submissions that are authored by external contributors or non-expert users, lack structure or have missing fields~\cite{fan2017where}. Consequently, these techniques can fall short in classifying \secissue.
\lili{I'm confused about the differences between text-mining techniques and ML techniques. Ain't ML also a form of text mining? ML techniques also use titles, descriptions, and comment threads of issues to mine them for their textual information and vectorize them for later classification. What are the differences?}\sogol{I edited the above and the below paragraph. Please double check.}
% from the rest of the submissions.
% text-mining alike techniques, that are dependent on structured textual data fall short in detecting \secissue which are itinerantly much vague and fragmented in discussion~\cite{burge2008rationale}.\sogol{Find another reference.} 

As a solution for the scarcity of structured information, researchers initiated the use of ML models on issue report summaries and unstructured text along with \textit{natural language processing (NLP)} techniques, such as regression analysis~\cite{fan2017where}, to extract additional features from them. For instance, Das et al. trained ML classifiers on probabilistic NLP features of a series of issue reports, that were previously attempted for classification only using frequency-based features~\cite{peters209textfiltering}, to achieve a much higher classification accuracy. In another study conducted by Gosteva-Popstojanova et al.~\cite{goseva2018identification}, researchers combined ML classifiers with varying types of feature vectors to detect security-related issues of four NASA datasets. 

Although demonstrating enhanced performance compared to solely vectorizing issues' structured information, these classifiers also have their own limitations. Specifically, ML models are known to rely on lexical cues, memorizing certain terminology and using them as spurious features in language inference related tasks such as issue report classification~\cite{gururangan-etal-2018-annotation, izmailov2022feature}. In other words, instead of learning the true context of the issue reports, ML models learn and use lexical shortcuts to predict labels, resulting in their failure when they attempt to classify more complex \secissue~\cite{mccoy-etal-2019-right}.
% IF trained with supervised algorithms, the ML models should be trained on the issue reports and their gold standard labels (i.e., ground truths). To address this problem, unsupervised training techniques have been introduced and utilized for training ML models for classification of issue reports~\cite{goseva2018identification}. Meanwhile, several research have demonstrated the benefits of supervised training, and teaching these ML models the ground truth of the data, specifically in classification tasks~\cite{}.\sogol{Find resources for this claim}.
% % Regardless of the used training algorithm, a huge dataset of issue reports is necessary to train these ML models.

Recent advancements in training autoregressive \textit{large language models (LLMs)}~\cite{min2023gptsurvey} and their enhanced performance in parsing, interpreting, and generating natural language, have their applications to be studied more extensively in varying software engineering tasks~\cite{ma-etal-2023-chain} including issue report classification~\cite{aracena2024applying, heo2025study}. If \textit{prompted} with detailed instructions~\cite{white2023prompt}, few-shot examples~\cite{Brown2020fewshot}, and %any additional 
necessary context-related information, LLMs can perform these tasks reasonably well.
% achieve a somewhat desirable performance in conducting these tasks; 
However, even with recent advancements, LLMs are still showing the tendency to \textit{hallucinate}, generating results that are factually inaccurate, incomplete, or inconsistent with the instructions they have received. Thus, it is probable for the LLMs to miss or incorrectly classify \secissue. Additionally, based on the model, using an LLM for classifying issues can take several hours and/or cost up to hundreds of dollars. Due to their limitations, triage engineers cannot use 
% either of 
the discussed detection tools to confidently identify \secissue in real-time. If such an issue is neglected, the underlying vulnerability propagates to dependent products which, in turn, jeopardizes stakeholder benefits~\cite{feng2014security, perwej2021systematic}.
\lili{We could also add a limitation: LLMs are costly in terms of time and money? If this is still the case}\sogol{Done!}
% till here

In the current paper, we propose \textbf{\classifier}, a framework to train transformer-based \textit{deep neural networks (DNNs)}, such as \textit{bidirectional encoder representations from transformers (\underline{\textbf{\bert}})}~\cite{devlin2019bert}, as effective \textit{masked language model (\mlm)} classifiers that can be used %confidently used by triage engineers 
in real-time to monitor issue trackers for \underline{\textbf{SE}}curity-related \underline{\textbf{IS}}sue submissions. Instead of training \dnns on ground truth labels of issues (i.e., security- and non-security-related), \classifier fine-tunes them on a list of keywords, that we call \keywords, that could semantically replace the label names if they appeared in issue reports. To ensure that the models are not memorizing lexical shortcuts to harvest spurious correlations with prediction labels, we mask the instances of \keywords so that the training can focus on the semantics embedded in context. We compare the performance of \dnns trained using \classifier with \baselinecount state-of-the-art issue report classifiers, including two ML- and three LLM-based ones, in detecting \secissue of a corpus of 10,000 GitHub submissions. 

Our evaluation results demonstrate that training with our framework is very\lili{externally? I'm not sure what this means here}\sogol{This was a typo :")} effective, with a \classifier-trained \dnn achieving 0.9849, 0.9924, and 0.9880 for \textit{precision}, \textit{recall}, and \textit{F1-score} performance metrics, surpassing ML-based baselines by 14.44\%-96.98\%, 15.40\%-93.07\%, and 14.90\%-94.72\%, accordingly. %over the ML-based baselines. %in detecting \secissue.The \classifier-trained \dnn 
Our classifier also outperforms the LLM-based baselines by 23.20\%-63.71\%, 36.68\%-85.63\%, and 39.49\%-74.53\% across precision, recall, and F1-score, respectively. To evaluate whether \classifier-trained \dnns can be used %by triage engineers 
as ready-made security-related issue detectors, we calculate the performance of our best-performing fine-tuned model on 1,000 totally unseen, \wild GitHub issue reports. Our analysis demonstrates that \classifier{}-trained models are capable of confidently detecting just-submitted \secissue, with 0.7123, 0.6860, and 0.6760 for precision, recall, and F1-score, comparable to those of prompting LLMs. \\
The contributions of our paper are outlined as follows.

\begin{itemize}
    \item[$\star$] We introduce a systematic procedure to identify \keywords, a list of semantically replaceable keywords with prediction label occurrences in issue reports. 
    \lili{We may also say that we systematically identify the \keywords...or introduce a systematic procedure to identify the \keywords?}\sogol{Done!}
    \item[$\star$] We design \classifier, a framework for training \dnns as capable and confident \secissuesingle classifiers that have minimal dependency on lexical cues. 
    \item[$\star$] We conduct extensive analyses, demonstrating \classifier{}'s effectiveness in training accurate, robust, and generalizable \secissuesingle classifiers.%~\footnote{To enable reproducibility, \classifier code and analysis scripts are available at \url{}.}
\end{itemize}

\lili{While the previous statements are correct describing how \classifier, they cannot intuitively explain why it can learn the context/semantic instead of the lexical shortcuts. It would be nice if we can intuitively and briefly introduce that. Something like: ``we mask the instances of the semantic surrogates so that the training can focus on the semantic embedded in the contexts...'' Something like this! Also, I'm not sure why we want to call the keywords as pseudo labels? I think in the context of our motivation, they may be better called lexical shortcuts? lexical shortcut keywords?}\sogol{I edited the above paragraph to reflect the intuition of how models do not memorize lexical cues. I also removed the \labels here because you are right this can get confusion in a concise description. I have clearly discussed in the methodology what we mean by \labels. Also, regarding the name change from \keywords to something like lexical cues, I think it is better not to do. lexical cues has a negative meaning often tied with the model memorizing certain terms and not learning the context, which is in contrast with what we are trying to do in this training. also, the \keywords are really terms that are semantically equivalent to the label names in the context of issue reports we have. They are not synonyms to the label names but in the context of issue reports we have excavated they have the same semantic meaning with the label names. So, I think semantic surrogates is an appropriate name, implying that they are not synonyms with the label names but they can carry the same semantic meaning (the implication of the surrogate).}
% a considerably higher detection accuracy compared to the ML-based baselines 
\lili{What are the baselines? We may want to discuss different baselines separately or discuss the baselines more in a more concrete way. e.g., it outperforms state-of-the-art xx by xx\%} \sogol{Done!}

%% file: 02_background.tex
\section{Background}\label{sec:background}
In this section, we introduce \dnns, discuss pre-trained language models, and briefly describe \mlm training.
% In this section we introduce \textit{Deep Neural Networks (\dnns)}, discuss pre-trained language models, and briefly describe \textit{Masked Language Model (\mlm)} training.

\subsection{Deep Neural Networks}\label{sec:dnn}
\textit{Deep neural networks (\dnns)} are non-linear ML models composed of multiple layers with numerous learnable parameters~\cite{samek2021deep}. %which, in turn, equip the model with impressive decision-making abilities~\cite{samek2021deep}. 
Each layer in a \dnn first applies a transformation on a collection of parameter values, including either all or a subset of the parameters of that layer (i.e., dense or mixture of experts (MoE) architectures)~\cite{vaswani2017attention}. The transformation is then followed by an element-wise nonlinearity. The collection of these layers creates a complex, multiscale, distributed architecture which, in turn, significantly enhances the decision making strategy of the model, leading to impressive performances in classification tasks, in addition to robustness to adversarial perturbation of input data and skewed data distributions~\cite{samek2021deep}.

\subsection{Pre-trained Language Models}\label{sec:pre-trained-lm}

A considerable population of \dnns has emerged as language models, pre-trained on vast textual corpora such as encyclopedias, enabling them to learn generic language features~\cite{tenney2019you}. These pre-trained \dnns can be further adapted to downstream tasks through two main strategies. In the former, the pre-trained architecture is integrated with an auxiliary, task-specific network to enhance task alignment. In the latter, a minimal set of new parameters is incorporated to the original architecture, followed by fine-tuning either the entire network or a subset of its parameters. A series of language models are unidirectional, such as \openai{}'s \textit{generative pre-trained transformers (\gpt)}~\cite{radford2018gpt}, capturing the relationships between words across the input text only from one direction and, in turn, generating the output autoregressively, token-by-token and from left-to-right~\cite{devlin2019bert}. Consequently, for tasks involving sentence inference or long-range comprehension, including issue classification, these models demonstrate a suboptimal performance~\cite{devlin2019bert}. In contrast, bidirectional transformers leverage context from both directions, achieving a higher classification performance over their unidirectional counterparts, particularly if they are fine-tuned on single tokens of input~\cite{devlin2019bert}.

\subsection{Masked Language Model Training}\label{sec:mlm}

Google's \textit{bidirectional encoder representations from transformers (\bert)}~\cite{devlin2019bert} is arguably one of the preeminent and widely adopted bidirectional transformer architectures for text classification tasks~\cite{meng2020text, bharadwaj2022github, siddiq2022bert, nozza2020mask}. \bert capitalizes on \textit{masked language model (\mlm)} training %~\cite{taylor1953cloze}
to incorporate the input textual context from both directions. \mlm randomly chooses and masks some tokens from the input and trains to predict those tokens from global dependencies of the surrounding context only using attention mechanisms~\cite{vaswani2017attention, nozza2020mask}. In other words, \bert learns a word's meaning not as is, but in the context of the semantics the word is used in. Due to its exceptional representation power that captures the long-range semantic dependencies of input texts, the application of \bert \mlm has been investigated across varying tasks from sentiment analysis~\cite{tian-etal-2020-skep}, to stance detection~\cite{kawintiranon2021knowledge}, multi-class text classification~\cite{meng2020text}, and log anomaly detection~\cite{lee2023lanobert}. Similarly, we also choose \bert to fine-tune it as an \mlm on the downstream task of detecting \secissue. 

%% file: 03_study_design.tex
\section{Methodology}\label{sec:methodology}
\begin{figure*}[tb]
    \centering
    % \footnotesize
    % \scalebox{1}{
    \includegraphics[width=0.99\linewidth]{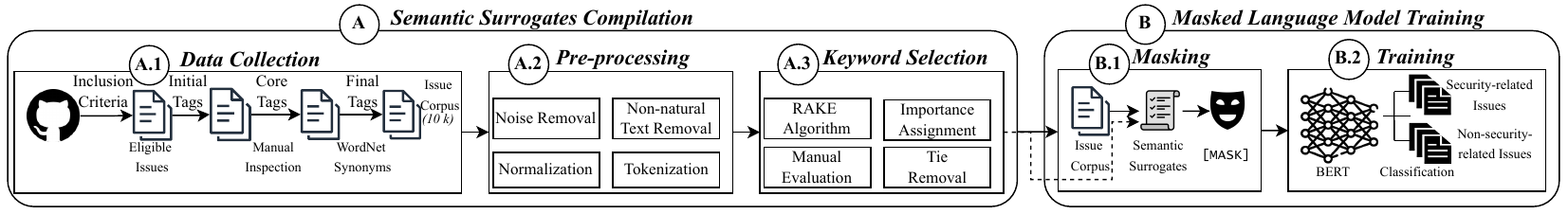}
    % }
    \caption{The overview of \classifier training framework}
    \label{fig:study-design}
\end{figure*}

We introduce \classifier, a novel framework to train \dnns for automatically detecting \secissue. Specifically, \classifier trains bidirectional transformer architectures as \mlms, i.e., fine-tuning their parameters when occurrences of certain keywords are masked in the corpus to prevent the models memorizing lexical shortcuts. We choose these keywords such that they can correctly replace label names (i.e., security- and non-security-related) should they appear in the issues. In other words, we compile the list of keywords as semantically equivalent vocabulary to prediction labels, calling them \keywords.
% , and use them as \labels of issue reports when \dnns are trained on them. 
Figure~\ref{fig:study-design} illustrates an overview of \classifier, consisting of \textit{\keywords Compilation}~\circled{A} and \textit{Masked Language Model Training}~\circled{B}. We describe each step in detail below. 
\input{031_classifier_data}
\input{032_classifier_design}

%% file: 031_classifier_data.tex
\subsection{\keywords Compilation~\circled{A}}\label{sec:security-surrogates}
For compiling the \keywords, after collecting a corpus of security- and non-security-related issue reports~\circled{A.1}, we pre-process the submissions to prepare them for keyword extraction~\circled{A.2}. Finally, we extract and rank keywords such that they can semantically represent issue label names~\circled{A.3}.

\subsubsection{Data Collection~\circled{A.1}}\label{sec:data-collection}

To identify the vocabulary that semantically represents the issue label names (i.e., security- and non-security-related), we first need to prepare a balanced corpus containing %a sufficient number of 
enough instances from both categories. To do so, we have initially targeted a set of 10 popular repositories, such as \textit{Microsoft vscode} and \textit{tensorflow}, to excavate their issues. %For example, repositories such as \textit{microsoft/vscode} and \textit{tensorflow/tensorflow} are included in this initial set. 
An issue with at least one relevant tag such as ``security'' or ``vulnerability'' (from hereon referred to as \textit{security-tags}) is labeled as security-related, whereas an issue without any security-tags is labeled as non-security-related. 

Guided by prior studies~\cite{chen2021let, zou2019branch, mastropaolo2023robustness}, we apply conservative project-level inclusion criteria to narrow down the initial population of selected repositories to a series of well-developed and well-maintained projects: $\geq 1{,}000$ stars; active within the past year; non-fork; a substantive closed issues history; $\geq 300$ commits; $\geq 50$ contributors; and $\geq 1$ merged pull request. These thresholds are intended to favor mature, well-maintained projects which, in turn, ensures the quality and generalizability of our issue report corpus. 
However, we have quickly realized that even well-known and well-maintained projects do not yield enough security-related issue reports. Security is relatively rare as a topic on issue trackers and security-tags are often missing, which further suppresses recall. 

Next, we have attempted to keep the above project-level filters while collecting security-related issues across all qualifying repositories, rather than pre-selecting and filtering within a small set of popular projects. This process also results in too few instances with security-tags, especially from large repositories where security-tagged issues remain sparse. Given this, we have decided to apply the inclusion restrictions at the issue level. %remove project-level restrictions and filter at the issue level instead. 
Specifically, we retain issue reports that are not pull requests, have nonempty titles and descriptions, and are submitted between 2022-01-01 and 2024-03-01 so that the dataset reflects contemporary security concerns. This shift preserves breadth while enforcing quality, 
% and consistency, 
yielding a larger and more representative collection of security-related issue reports. 
% \subsubsection{Label Collection}\label{sec:label-collection}

Because GitHub tags are author-defined and often inconsistent across projects and even across issue reports of the same repository, we treat them as a preliminarily signal rather than ground truth label names. 
% labeling as a semantic signal rather than a ground truth tag. 
Inspired from prior work on vulnerability disclosure by Kancharoendee et al.~\cite{kancharoendee2025categorizing}, we construct our initial set of security-tags to consist of
% the following five terms: 
``security'', ``vulnerability'', ``risk'', ``common vulnerabilities and exposures (CVE)'', and ``common weakness enumeration (CWE)''. CVE refers to a standardized catalog that assigns unique identifiers to publicly disclosed software vulnerabilities~\cite{mitreCVE}. CWE complements CVE by categorizing recurring design patterns and implementation flaws that lead to vulnerabilities~\cite{mitreCWE}.

We then conduct a round of systematic manual inspection of 250 GitHub issue reports drawn from a wide range of repositories and examine them for their tags that explicitly or implicitly indicate security relevance, considering both their linguistic meaning and their contextual use. From here, we identify four additional security-tags: ``common vulnerability scoring system (CVSS)'', ``CVSS/high'', ``CVSS/medium'', and ``CVSS/low'', where CVSS denotes a structured method for rating the severity of software vulnerabilities~\cite{firstCVSS}.
To broaden coverage while maintaining precision, we expand this core set of security-tags with \textit{WordNet}~\cite{wordnet_princeton}, a large lexical database of English vocabulary that groups words into sets of synonyms and captures their semantical relationship. Starting from our core set of security-tags, we retrieve their synonyms and related terms, then manually vet the candidates, keeping only those that are clearly and consistently tied to security. Following this process, we identify four other security-tags: ``exposure'', ``risk'', ``secure'', and ``vulnerable''. 

To extend our set of security-tags, we also experiment with \textit{Word2Vec}~\cite{word2vec_google_code}, which is a technique that receives textual inputs and generates numerical vector representations of terms that have a high semantic similarity with the input. %produces words as vectors based on their semantical similarity. 
Specifically, we use Word2Vec to identify the words that are semantically similar to our core set of security-tags. 
% compute cosine similarity between label embeddings and the embedding of \textit{``security''}, as used in recent work. 
% However, in our setting, \textit{Word2Vec} 
Word2Vec largely reproduces the synonyms that are already identified from WordNet and does not yield any additional meaningfully distinct security-tags; therefore, we do not adopt it. We also consider conducting \textit{snowball sampling}, an iterative procedure that expands the set of security-tags by incorporating new tags that frequently co-occur with existing seed terms~\cite{goodman1961snowball}. However, we decide against adopting this technique since first it risks propagating semantic noise which, in turn, can lead to topic drift, and second it inherently suffers from sample bias which can result in similar samples repeatedly reinforcing one another. Following the above inclusion criteria and our final set of security-tags, we create a balanced corpus of 10,000 GitHub issue reports consisting of 500 security- and 500 non-security-related issues. 
% Using the resulting security-label set, we classify an issue as security-related if it contains at least one of these labels; issues with none of them are treated as non-security. We apply the same dataset-level inclusion criteria described earlier to both groups. Our finalized list of GitHub security-related labels includes: \textit{“CVE"}, \textit{“CWE"}, \textit{“exposure"}, \textit{“risk"}, \textit{“secure"}, \textit{“security”}, \textit{“vulnerability"}, \textit{“vulnerable"}, \textit{“cvss/medium"}, \textit{“cvss/high"}, \textit{“cvss/low"}, \textit{“cvss"}.
\subsubsection{Pre-processing~\circled{A.2}}\label{sec:pre-processing}

% We next pre-process all security-related issue reports. 
Next, we pre-process our corpus. Following prior work on text pre-processing and software engineering text analysis, we (1) remove structural noise including Markdown headings and templates, HTML tags, checklists, file system paths, and raw URLs (removing any embedded hex codes while retaining their standalone meaningful terms)~\cite{chai2023comparison}; (2) normalize the text by lowercasing, removing numbers, special characters, and single-character tokens, and collapsing whitespace~\cite{tabassum2020survey}; (3) filter out non-natural-language content to down-weight code and logs~\cite{mantyla2018natural}; and (4) tokenize, apply part-of-speech-aware lemmatization, and remove English stop words~\cite{qiang2024does}. 
% We apply the same pre-processing step for non-security-related issue reports. %corpus to enable fair comparison.

\subsubsection{Keyword Selection~\circled{A.3}}\label{sec:keyword-selection}

For the pre-processed security- and non-security-related issue reports, we run the \textit{rapid automatic keyword extraction (RAKE)} algorithm~\cite{rose2010automatic}, which extracts key phrases from a body of text by ordering the importance of the terms based on their frequencies and co-occurrence relationships with the rest of vocabulary. For each issue report category, RAKE produces an ordered list of most important keywords, with their importance score, capturing the semantics of the corresponding context. We then manually evaluate both lists through iterative discussion sessions among three of the authors (the first, the second, and the corresponding authors) to remove keywords that are not meaningfully associated with their respective categories. The discussions continue until all inspectors confirm the relevance of emerged keyword lists. %for the issue report categories. 
For keywords appearing in both classes, we assign the term to the category with the higher importance score (i.e., the category where the keyword appears more frequently); ties are discarded. Finally, we select the top 50 keywords from each category to form the list of \keywords for that category.~\footnote{\keywords are available in the replication package.} \keywords serve as concise, interpretable proxies for security- and non-security-related issue content in our training framework. 

%% file: 032_classifier_design.tex
\subsection{Masked Language Model Training~\circled{B}~}\label{sec:mlm-training}

Supervised algorithms train \dnns to learn a mapping between the input data and their corresponding ground truth labels. The training enables the model to correctly infer the context of unseen inputs and predict their labels accordingly. However, such training conditions the \dnns on the exact vocabulary used in the training data. Consequently, the model memorizes specific lexical cues to rely on them as decision shortcuts when making predictions. In other words, the presence of certain terms in an unseen test input would suffice for the model to predict a label, without truly understanding the underlying context. %and making an informed decision based on it. 
To prevent \classifier{}-trained \dnns memorizing lexical shortcuts, rather than learning security-related context of issue reports, we set our framework to train the \dnns as \mlms over masked \keywords. %For this purpose, instead of training the \dnns on ground truth labels of the issue report we train them on any occurrences of \keywords in the corpus as \labels. 

Specifically, we analyze each issue report of our corpus to identify whether it contains any keywords from the list of \keywords for the category the issue belongs to. If such keyword exists, we assign it with the issue category as its pseudo-label. Training the \dnn on these keywords and their \labels would capture security-related semantics embedded in issue report context. %We then train the \dnn on each keyword and its surrounding context with pseudo-labels as correct predictions. As such, we ensure the model learns the semantics of issue reports. 
However, words are contextualized entities, meaning that a word can be interpreted differently based on the context it is used in~\cite{nozza2020mask}. In other words, other than a linguistic meaning, (i.e., the usual interpretation), the pragmatic interpretation of words depends on their surrounding vocabulary co-occurrence relationships. For instance, consider the example discussed by Meng et al.~\cite{meng2020text}. In a general context, the term ``exercise'' is interpreted as physical activity. However, in a sentence such as \texttt{``She was exercising her rights by voting.''}, ``exercise'' has a different pragmatic meaning from what is often inferred when the term is used. Thus, to train a \dnn as a classifier that is truly independent from lexical cues, not only the model should learn the security-related semantics of issue reports; but it should also learn the corresponding vocabulary pragmatics.

For this purpose, following the technique proposed by Meng et al.~\cite{meng2020text}, we hide any occurrences of \keywords in the corpus, by replacing them with the special \textsf{[MASK]} token~\circled{B.1}, before training the \dnn as an \mlm on their \labels~\circled{B.2}. 
% As the \dnn is trained to predict the issue categories over \textsf{[MASK]} tokens, 
% for determining the issue category the hidden semantic surrogate belongs to, the model is trained as an \mlm~\circled{B.2}. 
Similar to~\cite{meng2020text}, we also train a \bert model with an additional linear classification layer, to minimize the cross-entropy loss function. 
% to train our classifier, we fine-tune the parameters of the \bert architecture, with an additional linear classification layer, to minimize a cross-entropy loss function. 
Based on preliminary analyses, we fine-tune every learnable parameter.
% s in our classifier's final architecture.

% talk about the LLM to capture the pragmatics of the vocabulary Vs. the semantic capture of the BERT as an of-the-shelve classifer or when only used to predicit on the [CLS] token.

%% file: 04_experiment_design.tex
\section{Experiment Design}\label{sec:experiment-design}

To evaluate the effectiveness of \classifier in training \dnns as \secissuesingle classifiers, capable of learning context without memorizing lexical cues, we design a series of experiments investigating the following research questions.%\sogol{Think about whether the learning context and not memorizing lexical cues is necessary here.}

\reqone 
To respond, we breakdown the question into three parts. 

\reqoneone\\
\noindent \textbf{\underline{Motivation}:} As a measure of the effectiveness of \classifier in training \dnns as \secissuesingle classifiers, we calculate the best performance our fine-tuned \bert \mlm can achieve in detecting \secissue.\\
\noindent \textbf{\underline{Experiment Settings}:} We conduct training under a \textit{cross-validation (CV)}\cite{kim2009estimating} setting, splitting our corpus of issue reports into ten folds (i.e., 10-fold CV), where each fold is held out once as the validation set while the \dnn is fine-tuned on the remaining nine folds. We analyze the classification performance of the fine-tuned model for each fold by calculating \textit{precision}, \textit{recall}, and \textit{F1-score}. We report the overall performance of the \classifier-trained model as the statistical mean of these metrics across the ten folds. We use an NVIDIA T4 hardware accelerator with 15 GB of memory to execute the training pipeline. On average, the end-to-end training process, with a batch size of 32 and two worker threads, takes approximately 14 hours to complete.

\reqonetwo\\
\noindent \textbf{\underline{Motivation}:} Similar to any other training framework, exploring different \hyper settings to identify their optimal values is crucial for achieving the best classification performance. Thus, we are also set to find the optimal \hyper settings for \classifier.\\
\noindent \textbf{\underline{Experiment Settings}:} We evaluate whether increasing the number of times a \dnn observers a training sample, affects its classification performance. For this purpose, we conduct the 10-fold CV across increasing counts of epochs. %to analyze whether increasing their counts, improves the classification performance. 
To identify the appropriate epoch range for our experiment, we compare the classification performance of \bert \mlm when trained on a random 1,000 subset of our corpus against training on the full corpus. Based on prior recommendations~\cite{sajjad2023effect}, we set the optimal epochs count for training on the 1,000 data subset as three, resulting in 71\% classification accuracy. Meanwhile, training on the full corpus with only three epochs yields a considerably less accuracy of 60\%. Thus, we set the range of our training epochs to start from four and increase until no significant change for the classification performance is observed. We set other hyper-parameters including the learning rate the same as the original setting discussed for \bert~\cite{devlin2019bert}.

\reqonethree\\
\noindent \textbf{\underline{Motivation}:} For truly investigating the effectiveness of \classifier in training \dnns as successful \secissuesingle classifiers, other than calculating the detection performance of the fine-tuned model, it is also necessary to evaluate its performance compared to other \sota issue classifiers. \\
\noindent \textbf{\underline{Experiment Settings}:} We setup an experiment to compare the performance of \classifier-trained classifier in detecting \secissue of our corpus against \baselinecount (two ML- and three LLM-based) \sota issue classifiers. Our first baseline, \adapt~\cite{adaptirc}, adds adapters to transformer architectures and fine-tunes their parameters for %on the downstream task of 
classifying GitHub issues as bugs, enhancement requests, or questions. Our second baseline, is \fast, a specialized linear ML model which uses rank constraints and fast loss approximations to detect \secissue~\cite{secbugfast}. 
% \fast is fine-tuned on a corpus of 45,940 issue reports of five major GitHub repositories of Chromium, Ambari, Camel, Derby, and Wicket.
% , to accurately detect \secissue. 
% We consider the former baseline as a general-purpose, ML-based issue classifier while the latter is specialized in detecting \secissue. 
% To enable fair comparison, 
For fair comparison, we evaluate these ML-based baseline classifiers using the same performance metrics as \classifier-trained \dnns.  

We also compare the performance of \classifier-trained \dnns in detecting \secissue with \sota LLMs. To account for %the confounding effect of 
varying LLM architectures, we choose different families of models (proprietary in comparison to open-source), with varying architectural complexity (dense in comparison to MoE), and of different sizes (hundreds of billions of pre-trained parameters in comparison to models with fewer parameter counts) to serve as LLM-based baseline classifiers. Specifically, we use \openai{}'s \gpt family of models, including \gpt-4o-2024-08-06 and \gpt-4.1-2025-04-14 (from hereon referred to as \gpt-4o and \gpt-4.1, respectively). These two LLMs are representatives of proprietary LLM-based baselines with MoE, large architectures with an estimation of more than 200 billion parameters. We specifically choose these two models, compared to the newer \gpt-5 variant, as they are known to be the best-performing, most intelligent, non-reasoning LLMs~\cite{openai_models_compare}. we choose \meta{}'s \llama-2-70B-chat (from hereon referred to as \llama) as the open-source LLM-based baseline with a dense, medium-size architecture.  

We prompt the LLMs with each issue report of the corpus and instruct them 
% to conduct a binary classification task and 
to decide
% whether the target issue report hints at a security-related topic or in other words, 
whether the issue is security-related. To ensure the effectiveness of our prompts, we conduct a series of prompt engineering tasks, including carefully handcrafting the prompts following the \textit{context manager} structure proposed by White et al.~\cite{white2023prompt} by providing detailed instructions to the model for achieving highest accuracy while performing a binary classification task in the context of issue reports. We then apply spelling and grammar checkers for increasing clarity and removing any possible ambiguities~\cite{kamath2024scope}. 

We set the LLMs with the same decoding parameters: temperature equal to 0.7, to enable response diversity while minimizing randomness~\cite{openai_temperature}, and context window limit to ten tokens, to enforce focused predictions, i.e., the LLMs are instructed to output one or zero for security- and non-security-related predictions, respectively. Regardless of the temperature value, LLM responses are non-deterministic by nature. This entails that an LLM's classification decision can change across different prompting attempts for the same issue report~\cite{team2022chatgpt, ouyang2023llm}. To account for this residual stochasticity in the LLM responses, % we calculate their detection performance as a measure of \textit{pass@k}~\cite{chen2021evaluating}, which is the count of issue reports LLMs classify correctly given \textit{k} prompting attempts per issue report. 
we prompt the models three times for each issue report and calculate their performances in detecting \secissue as the mean of precision, recall, and F1-score. 

We use \openai{}'s batch endpoint~\footnote{https://platform.openai.com/docs/guides/batch} to prompt \gpt models.
% baseline classifiers. 
Each round of prompting for the full corpus 
% experiments, involving the 10,000 issue reports in our dataset, 
takes %approximately 
34 minutes on average to complete. The experiments cost USD 15.35 in total, with a token count of 13.34 million for inputs and 45.50 thousands for outputs (i.e., 13.39 million tokens in total). \llama experiments take an average of 8 hours and 41 minutes to complete using four AWS NVIDIA L40S hardware accelerators, at a cost of USD 8.30 per hour of execution.

\reqtwo\\
\noindent \textbf{\underline{Motivation}:} \classifier trains a \dnn as an \mlm by fine-tuning its parameters on a series of masked keywords. We believe the choice of masked terms affects the details of the context the \dnn learns, and subsequently, affects its prediction performance. Thus, we mask the occurrences of a list of keywords that can semantically replace the prediction labels in issue reports (i.e., \keywords). %By replacing \keywords with \textsf{[MASK]} tokens and training the \dnn to predict their \labels correctly, the classifier is 
Such training ensures the classifier actually learns semantics and pragmatics of security-relevant context embedded in the issues and does not memorize lexical cues for predictions which, in turn, result in the misidentification of more complex issues. 
% We believe the choice of masked terms affects the details of the context the \dnn learns, and subsequently, affects its prediction performance. 
We are set to analyze the extent to which the choice of masked terms can affect the capabilities of a \classifier-trained classifier in detecting \secissue.\\
\noindent \textbf{\underline{Experiment Settings}:} We compile a list of \random to replace them with \textsf{[MASK]} instead of masking \keywords. We then use \classifier to train the \dnn as an \mlm over the occurrences of randomly masked tokens.
% and their \labels. 
To identify the new set of keywords for masking, for each issue category we randomly choose 50 terms from the collection of vocabulary we have extracted using the RAKE algorithm (see Section~\ref{sec:keyword-selection}). In doing so we ensure that (1) chosen \random for each category are mutually exclusive and (2) no vocabulary in the set of \random belongs to the previously compiled list of \keywords. Thus, the difference in classification performance between a \dnn trained with \textsf{[MASK]} tokens replacing \keywords and one trained on randomly masked terms reflects the true importance (or lack thereof) of the masking vocabulary.
% use the below for when describing the experiment setting.
% Hence, we experiment with different set of \labels, once set as \keywords and once set as a list of random vocabulary.

\reqthree\\
\noindent \textbf{\underline{Motivation}:} To evaluate how good is the performance of %whether a triage engineer can use 
a \classifier-trained \dnn as an off-the-shelf classifier in %for 
detecting just-submitted \secissue in real-time, we need to investigate the generalizability of \classifier.\\
\noindent \textbf{\underline{Experiment Settings}:} Following the workflow described in Sections~\ref{sec:data-collection}~and~\ref{sec:pre-processing}, we create a balanced, \wild dataset of 1,000 issue reports consisting of 500 security- and 500 non-security-related issues. The dataset is then used as a fully unseen test set to evaluate the detection performance of the best-performing \classifier-trained classifier from \justrqone on it, serving as a measure of our framework’s generalizability. Additionally, we compare the performance of our ready-made classifier against the LLM baselines when they are prompted to detect \secissue of the \wild dataset.

%% file: 04_results.tex
\section{Results}\label{sec:results}
In this section we calculate the effectiveness and the generalizability of \classifier in training \dnns as \secissuesingle classifiers. Additionally, we quantify the impact of the masking vocabulary on the efficacy of \classifier.

\subsection*{\reqone}\label{sec:rq1-results}
\noindent \textbf{\underline{Observation 1 (\justrqoneone)}:}
The results of the 10-fold CV training are shown in Table~\ref{tab:rq1}. As can be observed, \classifier is very effective in fine-tuning \dnns as \secissuesingle classifiers, achieving up to 0.9738-0.9849, 0.9897-0.9924, and 0.9814-0.9880 for precision, recall, and F1-score, respectively.

\input{tables/rq1}

\noindent \textbf{\underline{Observation 2 (\justrqonetwo)}:}
Table~\ref{tab:rq1} also demonstrates the effect of increasing training epochs on the detection performance of the \classifier-trained classifier. As can be seen, increasing the epoch counts, enhances the capabilities of the classifier in correctly detecting \secissue, with six epochs of training resulting in the highest detection performance across all performance metrics (i.e., 0.9849, 0.9924, and 0.9880 for precision, recall, and F1-score, respectively). Additionally, setting the epochs=6 results in
% yields the most consistent performance across executions, %least variation in the performance metrics value with a 
standrad deviation values of only 0.0067, 0.0031, and 0.0063 for precision, recall, and F1-score, respectively. These values are significantly lower than those obtained from training with fewer epochs, which averaged 0.0264, 0.0053, and 0.0106, indicating that six training epochs yields more stable detection performance. 

\noindent \textbf{\underline{Observation 3 (\justrqonethree)}:}
Tables~\ref{tab:rq1}~and~\ref{tab:rq1-us-vs-others} demonstrate the detection performance of the \classifier-trained classifier compared to those of the \sota baseline classifiers.
% As can be observed, 
For any training epochs, the \classifier-trained classifier outperforms all baselines %in detecting \secissue of the corpus 
across all performance metrics. Specifically, our most accurate classifier (i.e., epochs=6) outperforms \fast and \adapt by 14.44\%-96.98\%, 15.40\%-93.07\%, and 14.90\%-94.72\% for precision, recall, and F1-score, respectively. 
Compared to \gpt-4o, \gpt-4.1, and \llama, our \classifier-trained classifier achieves 39.49\%, 42.88\% and 74.53\% higher F1-score, respectively. Improvements 
% percentage 
for precision and recall are 23.20\%, 23.65\%, and 63.71\% and 36.68\%, 38.68\%, and 85.63\% across \gpt-4o, \gpt-4.1, and \llama, respectively. The mean percentage improvement of our \classifier-trained classifier across the ML- and LLM-based baseline classifiers is 38.41\%, 48.10\%, and 48.21\% for precision, recall, and F1-score, respectively.

% preci/avg: 0.7116
%recall/avg: 0.6701
% f1/avg.: 0.6666

% On average a \classifier-trained classifier surpasses \sota issue classifiers by 45.33\%, 53.80\%, and 52.75\% for precision, recall, and F1-score, respectively. 
\input{tables/rq1-us-vs-rest}

\noindent \textbf{\underline{Observation 4 (\justrqonethree)}:} Table~\ref{tab:rq1-us-vs-others} also demonstrates the performances of our baselines in detecting \secissue against each other. As can be seen, \adapt has the lowest detection rate with precision, recall, and F1-score of only 0.5, 0.5140, and 0.5074, on par with randomly predicting issue labels with an accuracy of 0.5. All LLM-based baselines collectively outperform \adapt a minimum and maximum of 11.57\% and 39.59\% for F1-score while neither of them can detect \secissue with a better detection performance than \fast. \fast demonstrates the best baseline performance with a precision, recall, and F1-score of 0.8606, 0.8600, and 0.8599, exceeding others with improvements ranging from 7.66\% to 72.12\%, 18.44\% to 67.32\%, and 21.40\% to 69.47\%, respectively. As expected, among LLM-based baselines, \gpt-4o demonstrates the best detection performance, surpassing \llama as the worst performing LLM by 32.88\%, 35.82\%, and 25.12\% in precision, recall, and F1-score, respectively. Meanwhile, the difference in detection capabilities between \gpt-4o and \gpt-4.1 is only 0.36\%, 1.47\%, 2.43\% for precision, recall, and F1-score, respectively.

\noindent \textbf{\underline{Discussion}:} As evident from prior observations, \classifier is very effective in fine-tuning \dnns as accurate \secissuesingle classifiers. As expected, and similar to several other training frameworks, the count of training epochs has a direct relationship with the classification performance of the fine-tuned \dnn. However, regardless of the training epoch counts, our \classifier-trained classifier consistently demonstrates a superior detection performance compared to either of ML- or LLM-based baseline classifiers.

\begin{myframe}[width=\linewidth, top=0pt,bottom=0pt,left=0pt,right=0pt,arc=0pt,auto outer arc]
\small{\textbf{\underline{\textit{\justrqone Summary}}:}} \classifier-trains models that significantly outperform LLMs and transformer-based \secissuesingle classifiers by an average of 38.41\%, 48.10\%, and 48.21\% for precision, recall, and F1-score, respectively.
\end{myframe}

\subsection*{\reqtwo}\label{sec:rq2-results}
\noindent \textbf{\underline{Observation 5}:} Other than the detection performance of the \dnn when it is trained on masked \keywords, Table~\ref{tab:rq1} also demonstrates the model's performance when \textsf{[MASK]} tokens are replacing a series of random vocabulary in the corpus. The results show that the detection performance of \classifier-trained classier declines when %instead of semantically relevant tokens, 
random terms are masked. %to train the \dnn over them. 
Specifically, when replacing \random with \textsf{[MASK]} tokens instead of masking \keywords, precision, recall, and F1-score decrease by 13.82, 4.88, and 9.55 percentage points, respectively. In other words, the choice of masking vocabulary has an impact of up to 38.27\%-46.15\% and an average of 43.03\% on the success rate of \classifier.\\
\noindent \textbf{\underline{Observation 6}:} To evaluate whether the choice of masking vocabulary meaningfully contributes to the effectiveness of \classifier, we assess if the decline of performance, from masking \random rather than masking \keywords, is statically significant. To do so, we first establish whether the differences between the paired metric values across the CV folds, depicted in Table~\ref{tab:rq1}, are normally distributed.
For this, we conduct the \textit{Shapiro-Wilk} statistical test~\cite{shapiro1965analysis}, with the threshold of significance set to 0.05. The test demonstrates that the pair-wise differences 
% across the CV folds 
for precision and F1-score do not follow a normal distribution (i.e., $p-value<0.05$) while differences for recall paired values are normally distributed (i.e., $p-value>0.05$).
Following these observations, to assess the significance of the decline across the performance metrics,
% metric paired values,
% across the CV folds, 
we conduct the non-parametric \textit{Wilcoxon signed-rank} statistical test~\cite{wilcoxon1992individual} for precision and F1-score and the parametric \textit{Paired T} statistical test~\cite{student1908probable} for recall.
% As we are conducting multiple comparisons, 
We also apply the \textit{Bonferroni} correction~\cite{armstrong2014Bonferroni} on the threshold of significance (i.e., $\alpha/k$ with $\alpha$ being the confidence limit=0.05 and k being the count of testing hypotheses which is three accounting for precision, recall, and F1-score). For both %Wilcoxon signed-rank and Paired T statistical 
tests, we correct the threshold of significance as $0.05/3=0.017$. The conducted statistical tests demonstrate that the detection performance of the \classifier-trained classifier significantly decreases across all three metrics 
% of precision, recall, and F1-score 
(i.e., $p-value<0.017$) when parameter fine-tuning is done over randomly masked tokens rather than masked \keywords.\\
\noindent \textbf{\underline{Observation 7}:} As evident by the results in Table~\ref{tab:rq1}, masking the random set of keywords also increases the standard deviation values across the performance metrics compared to when the \dnn is trained on the masked tokens of \keywords. Specifically, the ratios of sample variances across the performance metrics are substantially higher than the expected deviations, if the choice of masking vocabulary did not have a substantial impact on the detection performance of the \classifier-trained classifier~\cite{fisher1924distribution}. In other words, the classifier's performance variability significantly increases when masks are assigned randomly. This, in turn, suggests that the choice of masking vocabulary indeed has a substantial impact on the success-rate of the \classifier-trained classifier.\\
\noindent \textbf{\underline{Discussion}:} Our experiment demonstrates that the choice of vocabulary for replacement with \textsf{[MASK]} is important for a successful training with \classifier. Not only the detection performance of the \classifier-trained classifier declines when ad-hoc vocabulary are masked, the variability of its performance also increases significantly, leading to substantially fluctuating predictions for the same issue report.

\begin{myframe}[width=\linewidth, top=0pt,bottom=0pt,left=0pt,right=0pt,arc=0pt,auto outer arc]
\small{\textbf{\underline{\textit{\justrqtwo Summary}}:}} The success of \classifier derives from two aspects: (1) training for predicting \labels of masks, to prevent lexical cue memorization, and (2) using \keywords as masking vocabulary, to enhance prediction accuracy and stability.
\end{myframe}

\subsection{\reqthree}\label{sec:rq3-results}
\noindent \textbf{\underline{Observation 8}:} Table~\ref{tab:rq3-us-vs-others} demonstrates the detection performance of our best-performing \classifier-trained classifier (\textit{\justrqoneone} and \textit{\justrqonetwo}) in detecting \secissue of the fully unseen collection of 1,000 \wild issue reports. As can be observed, our classifier achieves a precision of 0.7123, demonstrating its confidence in correctly detecting \secissue that it has never encountered before. On other hand, the recall of our classifier is measured slightly lower and at 0.6860, indicating its conservative behavior in flagging issue reports that their prediction probabilities are marginally greater than the decision threshold of 0.5. \\ %is of  meaning that it classifies 31.40\% of \secissue as non-security-related.\\
\input{tables/rq3-us-vs-rest}
\noindent \textbf{\underline{Observation 9}:} Table~\ref{tab:rq3-us-vs-others} also demonstrates the performance of the \classifier-trained \dnn in classifying the \wild dataset in comparison to the LLM-based baselines. %As evident from the table, 
Our classifier performs almost on par with the LLMs when they are used as ready-made detectors for \secissue (i.e., 0.7123, 0.6860, and 0.6760 for precision, recall, and F1-score of our classifier compared to those across the LLMs, which averaged 0.7741, 0.7265, and 0.7336, respectively). This is particularly interesting due to the possibility of data leakage for LLMs, i.e., because of their training cutoff dates, the LLMs may have already observed the \wild issues during pre-training on open-source GitHub repositories. Nonetheless, our \classifier-trained classifier achieves comparable detection performance, despite the disadvantage of not having previously observed or learned from the \wild dataset.\\
\noindent \textbf{\underline{Discussion}:} A %triage engineer can use a previously 
\classifier-trained \dnn can be used as a ready-made classifier to detect recently submitted \secissue, achieving a performance comparable to prompting LLMs while avoiding their associated inference costs.

\begin{myframe}[width=\linewidth, top=0pt,bottom=0pt,left=0pt,right=0pt,arc=0pt,auto outer arc]
\small{\textbf{\underline{\textit{\justrqthree Summary}}:}} A \classifier-trained \dnn performs as cost-effective counterpart to LLMs when used as ready-made classifiers on fully unseen data, detecting \secissue with 0.7123, 0.6860, and 0.6760 precision, recall, and F1-score, respectively.
\end{myframe}

%% file: tables/rq1.tex
% \usepackage{booktabs}
% \usepackage{graphicx}
% \usepackage{multirow}
\begin{table*}[!htp]
\caption{The performance of \classifier-trained classifier across 10-fold CV}
\label{tab:rq1}
\centering
\footnotesize
% \resizebox{0.98\linewidth}{!}{%
\begin{tabular}{@{}ccrrr@{\hspace{1.5em}}rrr@{}}
\toprule
% two empty cells above the grouped headings: (labels) & (epochs)
& & \multicolumn{3}{c}{\textbf{Mean}} & \multicolumn{3}{c}{\textbf{Standard Deviation (Std.)}} \\ 
\cmidrule(lr){3-5}\cmidrule(lr){6-8}
\textit{Masks} & \textit{Epochs} & \textit{Precision} & \textit{Recall} & \textit{F1-Score} & \textit{Precision} & \textit{Recall} & \textit{F1-Score} \\ \midrule
\multirow{3}{*}{\shortstack{Semantic\\Surrogates}}
                 & 4               & 0.9738                                 & 0.9898                              & 0.9815                                & 0.0275                                 & 0.0057                              & 0.0114                                \\
                 & 5               & 0.9740                                 & 0.9897                              & 0.9814                                & 0.0252                                 & 0.0049                              & 0.0098                                \\
                 & \circled{6}               & \textbf{0.9849}                        & \textbf{0.9924}                     & \textbf{0.9880}                       & \textbf{0.0067}                        & \textbf{0.0031}                     & \textbf{0.0063}                       \\ \midrule
Random\\Keywords          & 6               & 0.8467                                 & 0.9436                              & 0.8925                                & 0.1099                                 & 0.0347                              & 0.0693                                \\ \bottomrule
\end{tabular}%
% }
\end{table*}

%% file: tables/rq1-us-vs-rest.tex
% \usepackage{booktabs}
% \usepackage{multirow}
% \usepackage{graphicx}
\begin{table}[!htp]
\centering
\caption{\centering The performance of \classifier-trained and baseline classifiers}\sogol{(1): 10-fold VC for \\lama, (2): \wild separate from 10-fold CV}
\label{tab:rq1-us-vs-others}
\centering
\footnotesize
\begin{tabular}{@{}p{1cm}cc@{\hspace{0.2cm}}rr@{\hspace{0.2cm}}r@{}}
\toprule
                                            & \textbf{Type}              & \textbf{Classifier}                                          & \textbf{Precision} & \textbf{Recall} & \textbf{F1-Score} \\ \midrule
\multirow{5}{*}{Baseline}                   & \multirow{2}{*}{ML-based}  & \adapt                                        & 0.5000             & 0.5140          & 0.5074            \\
                                            &                            & \fast                                     & 0.8606             & 0.8600          & 0.8599            \\ \cmidrule(l){2-6} 
                                            & \multirow{3}{*}{LLM-based} & \gpt-4o                                       & 0.7994             & 0.7261          & 0.7083            \\
                                            &                            & \gpt-4.1                                      & 0.7965             & 0.7156          & 0.6915            \\
                                            &                            & \llama                                        & 0.6016             & 0.5346          & 0.5661            \\ \midrule
\multirow{1}{*}{\classifier} 
                                            &      10-Fold CV                      & \bert \mlm                     & \textbf{0.9849}    & \textbf{0.9924} & \textbf{0.9880}   \\ \bottomrule
\end{tabular}
\end{table}

%% file: tables/rq3-us-vs-rest.tex
% \usepackage{booktabs}
% \usepackage{multirow}
% \usepackage{graphicx}
\begin{table}[!htp]
\centering
\caption{\centering The performance of ready-made \classifier-trained classifier and LLM-based baselines}\sogol{(1): 10-fold VC for \\lama, (2): \wild separate from 10-fold CV [d]}
\label{tab:rq3-us-vs-others}
\centering
\footnotesize
\begin{tabular}{@{}p{1cm}cc@{\hspace{0.2cm}}rr@{\hspace{0.2cm}}r@{}}
\toprule
                                            & \textbf{Type}              & \textbf{Classifier}                                          & \textbf{Precision} & \textbf{Recall} & \textbf{F1-Score} \\ \midrule
\multirow{3}{*}{Baseline}                   & \multirow{3}{*}{LLM-based} & \gpt-4o                                       & 0.8536             & 0.8100          & 0.8040            \\
                                            &                            & \gpt-4.1                                      & 0.8583             & 0.8258          & 0.8218            \\
                                            &                            & \llama                                        & 0.6104             & 0.5436          & 0.5751            \\ \midrule
\multirow{1}{*}{\classifier} 
                                            & In-the-Wild                & \bert \mlm                     &        0.7123            & 0.6860                 & 0.6760                      \\ \bottomrule
\end{tabular}
\end{table}

%% file: 05_implications.tex
\section{Practical Implications}\label{sec:pi}
In this section of the paper, we discuss the implications of our results for researchers and practitioners.

\subsection*{\imp 1: If fine-tuned properly, bidirectional transformers are much better issue classifiers than unidirectional architectures.}\label{sec:pi1}
\textit{Observations 1 (\justrqoneone) and 3 (\justrqonethree)} demonstrate that the \classifier-trained classifier outperforms LLM-based baselines in detecting security-related issues within the %in-distribution 
training corpus. \textit{Observations 8 and 9 (\justrqthree)} further show that the \classifier-trained classifier generalizes effectively to fully unseen issue reports, achieving comparable performance to those of the LLMs. %on out-of-distribution data. 
% Our classifier has not observed any of the \wild issues before while the LLM-based baselines have been possibly exposed to the issues during pre-training; yet, our classifier achieves comparable performance to those of the LLMs. 
This is specifically important as these LLMs are possibly exposed to \wild issues during pre-training while our \classifier-trained classifier has not previously observed any of them. \textit{Observation 4 (\justrqonethree)}, on the other hand, illustrates the inefficacy of \adapt compared to the LLM-based baselines in detecting \secissue. Similar to \classifier, \adapt also fine-tunes the parameters of bidirectional transformers to train them as issue classifiers. However, among all classifiers used in this study, \adapt performs the worst, essentially as a random guesser and making predictions by flipping a coin with 0.5 accuracy. In other words, the best- and the worst-performing \secissuesingle classifiers are both bidirectional transformers, but fine-tuned with different training frameworks. This finding highlights the potential of these architectures to outperform unidirectional LLMs as issue classifiers, if proper frameworks are adopted for their fine-tuning. \\
\noindent \textbf{\underline{To Developers}:} We recommend fine-tuning bidirectional transformers as ready-made detectors, instead of prompting LLMs, for monitoring issue trackers and assigning submissions to appropriate developer teams.

\subsection*{\imp 2: The choice of masking vocabulary during \mlm training significantly impacts the performance of the classifier.}\label{sec:pi2}
\textit{Observation 5 (\justrqtwo)} illustrates the decline in the detection performance of the \classifier-trained classifier when random vocabulary are replaced with \textsf{[MASK]}, rather than masking \keywords. Masking \random results in the decrease of all performance metrics, with precision demonstrating the most and recall showing the least decline (i.e., 13.82 and 4.88 percentage points, respectively). The statistical tests conducted in \textit{Observation 6 (\justrqtwo)} demonstrate that the decline in the detection performance %resulted from randomly masking the corpus, 
is statistically significant, further emphasizing the role of training over masked \keywords in compelling the model to base its predictions on non-spurious features. Additionally, \textit{Observation 7 (\justrqtwo)} reports the significant increase in the classifier's prediction variability, from training on masked \keywords to training on random masks.
These statistics highlight the impact of choosing context-aware vocabulary for masking during \classifier, or similar \mlm-based frameworks, for training not only accurate but also stable issue classifiers.

\noindent \textbf{\underline{To Researchers}:} To implement an effective \mlm-based training framework, we recommend compiling a list of semantically-related, context-aware keywords for masking. This approach enables the classifier to truly learn the details, semantics, and pragmatics of the context. 

\subsection*{\imp 3: Other than \textsf{[MASK]} tokens, the confidence of the classifier depends on \textsf{[CLS]}.}\makeatletter
\def\@currentlabel{VI-C}
\makeatother\label{sec:pi3}

\textit{Observations 6 and 7 (\justrqtwo)} discuss the impact of masking vocabulary on the accuracy and prediction variability of the \classifier-trained classifier. If masked tokens are selected randomly or are not semantically relevant to the context,
% and the categories of data,the classifier's detection capabilities substantially fluctuates across different splits of the same input. In other words, 
the classifier’s prediction of the same issue may vary across different attempts. This stochasticity of predictions in return, undermines the confidence of developers in using the \classifier-trained classifier as an effective automatic detection tool. Similarly, \textit{Observations 8 and 9 (\justrqthree)} illustrate the confidence of the \classifier-trained classifier as a ready-made detector in distinguishing \secissue in real-time. Although less precise than when fine-tuned on data characteristics, our classifier detects \secissue in real-time with precision that substantially surpasses %\adapt and 
\llama{}'s and is comparable to the average across all LLM-based baselines.
% \gpt-based baselines.  

\begin{figure}[htp]
  \centering
  % Reduce internal column padding if you want even tighter spacing:
  \setlength{\tabcolsep}{3pt}%
  \resizebox{\columnwidth}{!}{%
    \begin{tabular}{ccc}
      \subcaptionbox{Overall \label{fig:cm-total}}{%
        \includegraphics[width=4cm,height=4cm]{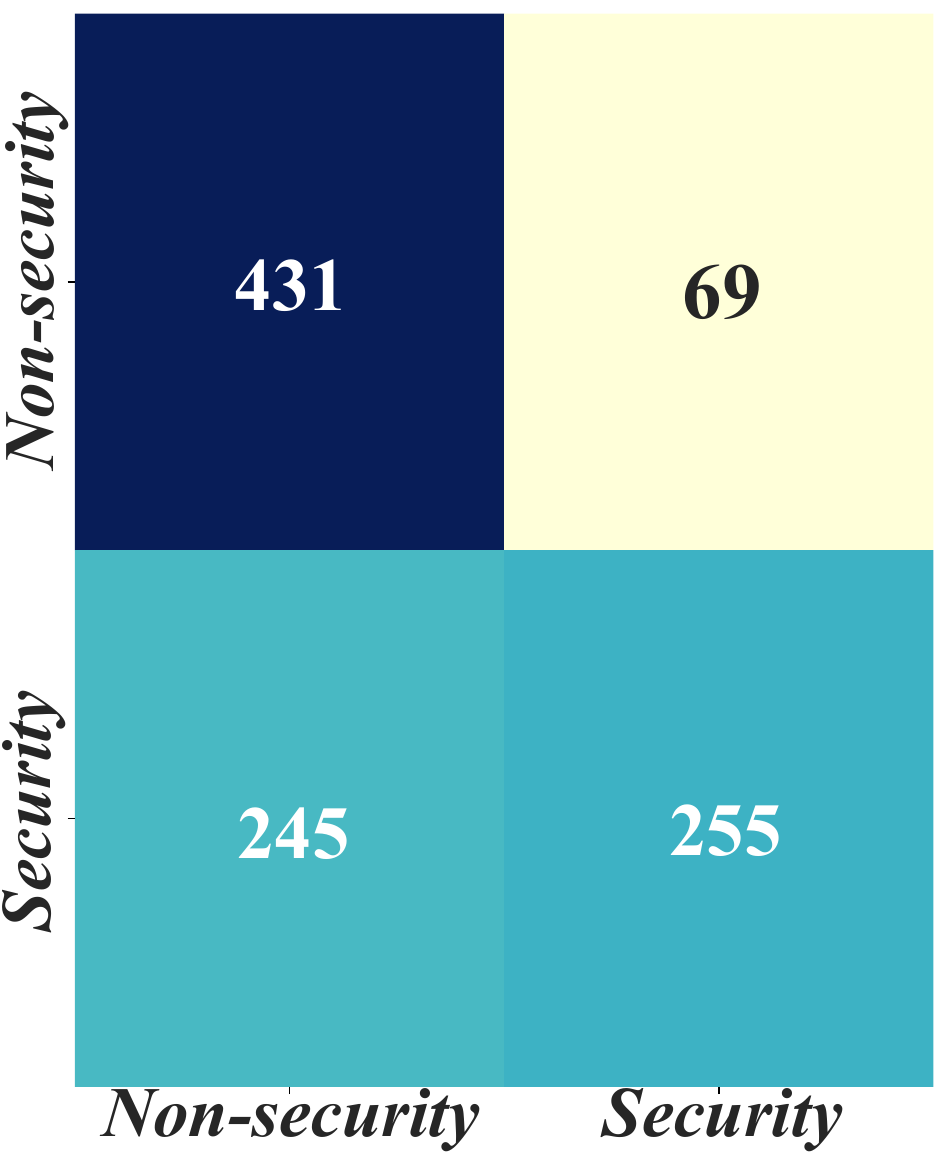}%
      } &
      \subcaptionbox{\textsf{[MASK]} predictions\label{fig:cm-masked}}{%
        \includegraphics[width=4cm,height=4cm]{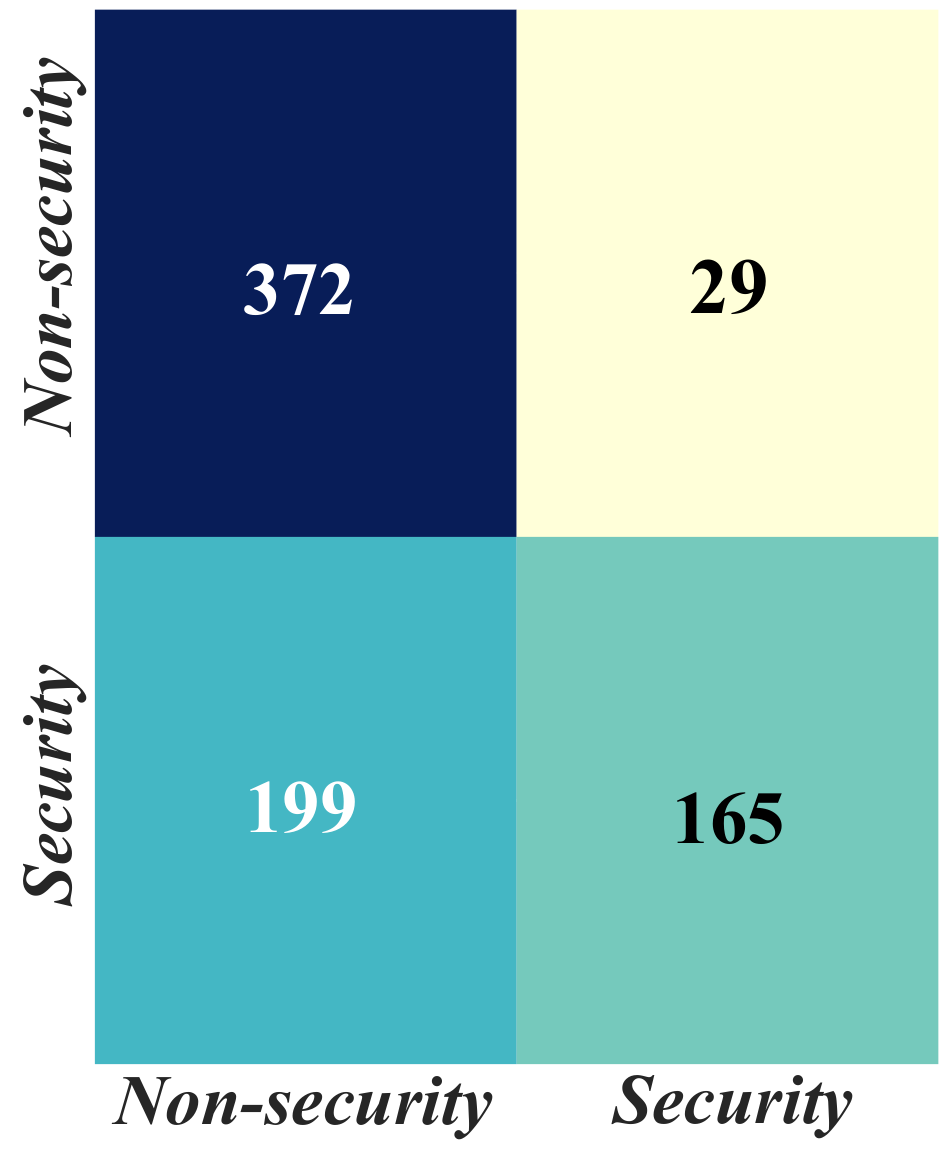}%
      } &
      \subcaptionbox{\textsf{[CLS]} predictions\label{fig:cm-cls}}{%
        \includegraphics[width=4cm,height=4cm]{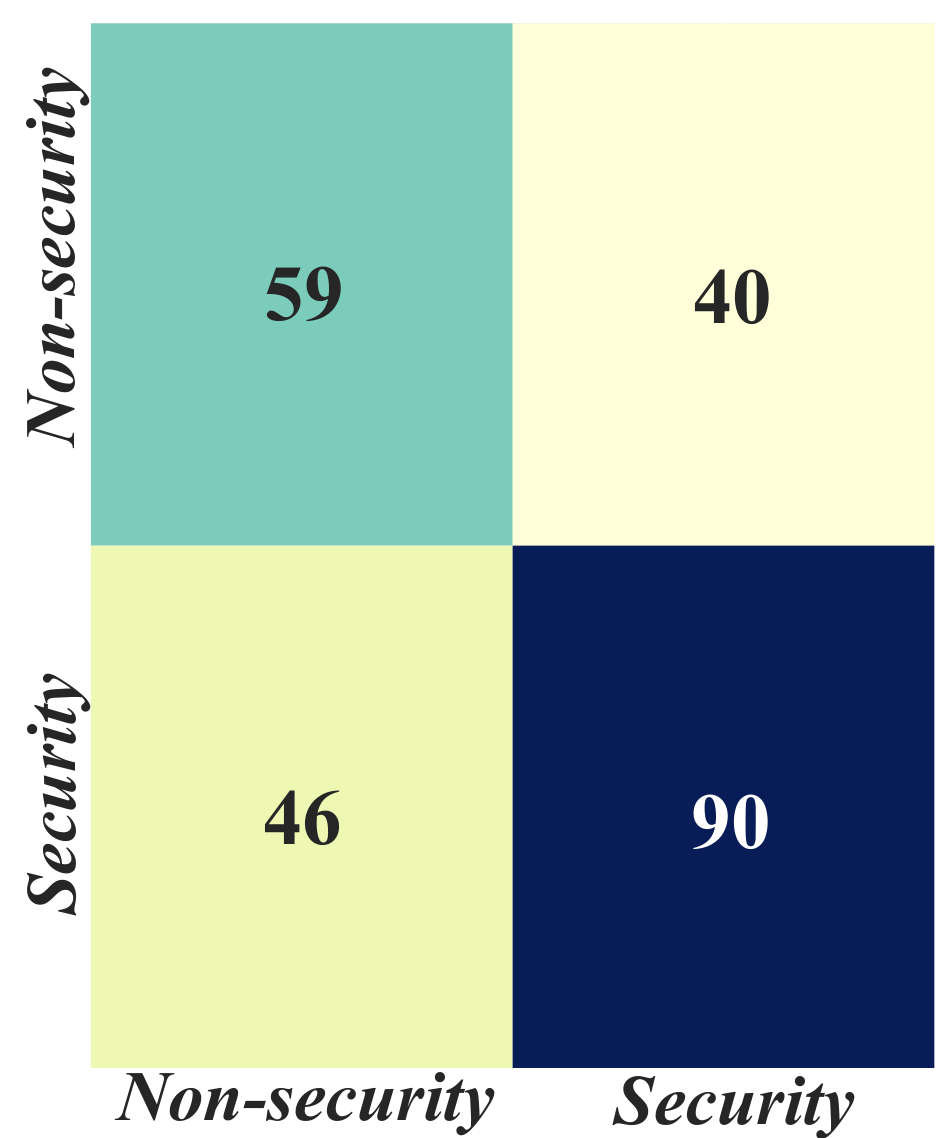}%
      }
    \end{tabular}%
  }% end resizebox
  \caption{The \wild\ CM for \classifier-trained classifier\\ \centering\small{X axis: predictions, Y axis: labels}}
  \label{fig:cm}
\end{figure}

To further investigate the underlying reasons for the decline of precision when our trained classifier is used on the \wild dataset, we calculate the count of correctly classified \secissue (i.e., \textit{True-Positives (TP)}) and the count of \textit{False-Positives (FP)} (i.e., false alarms for issues that report bugs with no security threats). Figure~\ref{fig:cm}, illustrates the \wild classification performance decomposition as \textit{Confusion Matrices (CM)}. Figure~\ref{fig:cm-total} illustrates the TPs and FPs across the full population of the \wild data. Figure~\ref{fig:cm-masked}, depicts the TPs and FPs across the \wild issues that contain at least one masked Semantic Surrogate while Figure~\ref{fig:cm-cls} shows the counts of TPs and FPs for the issues that have no occurrence of \keywords (i.e., zero masks). In the former setting, the ready-made classifier predicts an issue as either security- or non-security-related by conducting majority voting over the predictions for \labels of \textsf{[MASK]} tokens. For example, if a security-related issue is masked for three \keywords, and for two of those the predicted \labels are security-related, the majority of the classifier's casted votes detect the issue as security-related, making it a TP. In the latter, as issues do not contain any \keywords, there exists no \textsf{[MASK]} tokens for the classifier to predict their \labels and assign the prediction label for the issue category as their consensus. Consequently, the prediction decision falls back on the \textsf{[CLS]} token head (i.e., the standard classification layer in the architecture) similar to when the transformer is used as an off-the-shelf classifier~\cite{devlin2019bert}. As expected, the classifier's precision is much higher when it detects \secissue by assigning the majority of votes over \textsf{[MASK]} token predictions (i.e., precision=0.7464) compared to predictions over the \textsf{[CLS]} head (i.e., precision=0.6373). The same goes for the recall of the ready-made classifier %with its overall recall across the full population of the unseen data as 0.6860 and its recall 
over \textsf{[MASK]} and \textsf{[CLS]} tokens as 0.7020 and 0.6341, respectively. For both settings, the performance metrics are calculated as weighted measures due to the population imbalance between the security- and non-security-related issue reports.

% \begin{figure*}[htp]
%     \centering
%     \begin{subfigure}[b]{0.32\textwidth}
%         \centering
%         \includegraphics[height=4cm, width=4cm]{figures/edited_2025-10-01_16-47_ALL__sample_round_1_training_fold_2_trained_model.pdf}
%         \caption{Overall CM}
%         \label{fig:cm-total}
%     \end{subfigure}
%     \hfill
%     \begin{subfigure}[b]{0.32\textwidth}
%         \centering
%         \includegraphics[height=4cm, width=4cm]{figures/edited_2025-10-01_16-47_MLM__sample_round_1_training_fold_2_trained_model.pdf}
%         \caption{CM over \textsf{[MASK]} token predictions}
%         \label{fig:cm-masked}
%     \end{subfigure}
%     \hfill
%     \begin{subfigure}[b]{0.32\textwidth}
%         \centering
%         \includegraphics[height=4cm, width=4cm]{figures/edited_2025-10-01_16-47_CLS_sample_round_1_training_fold_2_trained_model.pdf}
%         \caption{CM over \textsf{[CLS]} token predictions}
%         \label{fig:cm-cls}
%     \end{subfigure}
%     \caption{\bert \mlm \wild classification decomposition}
%     \label{fig:cm}
% \end{figure*}

\noindent \textbf{\underline{To Researchers}:} To improve the performance of the \classifier-trained classifier as a ready-made detector, we recommend enabling majority voting over \textsf{[MASK]} token predictions for all issue reports. To do this, if \keywords are not present in all issues, a second list of semantically equivalent vocabulary should be compiled to represent the labels for only those reports with no occurrences of \keywords.

%% file: 06_threats_to_validity.tex
\section{Threats to the Validity}\label{sec:threats}
We discuss the threats to the validity of our study in a breakdown of construct, internal, and external.
\subsection{Construct Validity}\label{sec:construct-validity}
These threats pertain to the correctness of our measurements. To capture the effectiveness of \classifier in fine-tuning \dnns as \secissuesingle classifiers, we conduct 10-fold CV training, ensuring of no data leakage between the held-out validation fold and the remaining training folds. %used as the training set. 
Additionally, after the completion of training for each fold, fine-tuned parameters of the \dnn are reset to pre-training weights so the mean of performance metrics across the folds are correct representations of the detection capabilities of the \classifier-trained classifier. %across all the folds. 
Nonetheless, precision, recall, and F1-score %, regardless of their frequent use in prior similar studies, 
may not be proper choices for capturing the effectiveness of \classifier. %in fine-tuning \dnns as \secissuesingle classifiers. 
However, these metrics are frequently adopted in literature to calculate the performance of ML-based systems tailored for classifying issue reports~\cite{bharadwaj2022github, siddiq2022bert}.

\subsection{Internal Validity}\label{sec:internal-validity}
Inherited challenges constitute internal threats to the validity of our study. %To compile the list of \keywords for security-related issue reports, we first populated our corpus with security-tagged issue reports. 
To mine for \secissue, %identify all such issues, 
we first identify the GitHub tags that their meaning is closely related to security, extending the list in three folds: (1) adding security-tags that are identified in literature, (2) manually inspecting issues for security-tags that have been missing in our initial set, and (3) retrieving security-relevant synonyms for the core tags from WordNet. While we aim to compile a comprehensive set of plausible security-tags, our list may not include all tags that are currently used or will be assigned to \secissue in the future.
We follow a systematic procedure to compile \keywords as a comprehensive list of context-aware substitutes for issue report labels. Our experiments (i.e., \textit{\justrqtwo}) also demonstrate the effectiveness of fine-tuning the \dnns over masked \keywords rather than masking random keywords for creating a more accurate classifier with less prediction stochasticity. However, our compiled list of \keywords may not capture every possible security-related semantics, such as certain jargon that may have not appeared in our corpus of GitHub issue reports.

\subsection{External Validity}\label{sec: extrenal-validity}
External validity concerns the generalizability of our study. \classifier fine-tunes \bert as an \mlm, conditioning the \dnn to predict \secissue over masked \keywords. Fine-tuning as an \mlm is only possible, if a \dnn is trained with mask prediction objective during pre-training. This includes \bert and all its variants including but not limited to RoBERTa~\cite{liu2019roberta}, ALBERT~\cite{lan2020albert}, and DeBERTa~\cite{he2021deberta}. Additionally, any encoder-only transformer, such as multilingual \dnns with \mlm head in pre-training~\cite{lample2019xlm}, and encoder-decoder architectures with access to the \textsf{[MASK]} token~\cite{lewis2020bart} can be trained with \classifier as \secissuesingle classifiers. As shown by our results (\textit{\justrqthree}), in case of no Semantic Surrogate occurrence in issues, our \classifier-trained classifier makes predictions over the \textsf{[CLS]} head. However, as discussed in Section~\ref{sec:pi3}, to increase the confidence of the classifier, a secondary list of semantically equivalent vocabulary to labels can be compiled to only cater to issues whose 
% security-related 
context is not captured by \keywords.

%% file: 07_related_work.tex
\section{Related Work}\label{sec:related-work}
In this section of the paper, we briefly discuss the literature on automatic issue classification techniques, dividing them to ML- and LLM-based approaches.
\subsection{ML-based Issue Classification Techniques}\label{sec:ml-studies}
Peters et al.~\cite{peters2017text} processed issue reports of Chromium and four major Apache projects including Ambari, Camel, Derby, and Wicket to create a dataset of 45,940 issues with correct GitHub tags. They removed any non-security-related issue with security-tags to enhance the detection performance of classifiers. Their dataset, was used by Alqahtani~\cite{secbugfast} to fine-tune \fast. The same dataset was also used by Das and Rahman~\cite{das2018security} to evaluate the capabilities of ML models in detecting \secissue. Several other studies also employed ML models to classify issue reports. Zou et al.~\cite{zou2018automatically} combined the discriminative power of ML models with meta-features to classify 23,262 issue reports of three major projects: Firefox, Thunderbird, and Seamonkey as security- and non-security-related. In another study conducted by Kallis et al.~\cite{kallis2019tickettagger}, ML models are used for multi-class classification of GitHub issue reports. 
% Multi-class classification has also been done for issue submissions to Jira issue tracking system, classifying the reports into defects, improvements, or feature requests, through an NLP-based plugin~\cite{alhindi2023labeler}. 
Other than GitHub, ML models are also used for classifying issue submissions to Bugzilla and Jira issue trackers~\cite{wu2021data, alhindi2023labeler}.

\subsection{LLM-based Issue Classification Techniques}\label{sec:llm-studies}
Other than ML models, LLMs are also used for classifying issue reports. Aracena et al.~\cite{aracena2024applying} used \gpt-3.5 Turbo for classifying issues of five GitHub projects. Building upon the previous study, Heo et al.~\cite{heo2025study} investigated the benefits of prompt engineering in effectively instructing a varying range of LLMs, including two different \gpt variants and a \llama model, to classify GitHub issues. 
% into categories of bug, feature request, and questions. 
Du et al.~\cite{du2024llm} proposed \textsc{LLM-BRC}, a highly accurate LLM-based framework specialized for classifying issues of deep learning libraries, achieving an F1-score of 0.9875. Colavito et al.~\cite{colavito2024leveraging} compared the performance of \gpt-3.5 with two bidirectional transformer architectures, including RoBERTa, in categorizing issue reports into four different classes of bug, feature requests, questions, and documentation. Their results demonstrated that although \gpt-3.5 can achieve a relatively high accuracy, fine-tuned \dnns still outperform the LLM in the issue classification task. 

\vspace{5pt}
Above studies mainly focus on detecting bugs from other submissions. Ones that target only \secissue often suffer from suboptimal performance, limited generalizability across projects, or inability to classify newly submitted issues in real-time. In contrast, our framework trains classifiers independent of lexical cues or project-specific issue reports. Thus, a \classifier-trained classifier can generalize across projects and serve as a ready-made \secissuesingle detector. 
% \sogol{[METHODOLOGY]. Add the benefits of using issue reports instead of issues of certain projects to ensure the classifier can have inter-project generalizability.}

% \sogol{What are the challenges they faced in their work that our technique is addressing.}

%% file: 08_conclusion.tex
\section{Conclusion}\label{sec:conclusion}
In this paper, we introduce \classifier, a framework for producing \secissuesingle classifiers. \classifier trains a bidirectional transformer as an \mlm by (1) effectively masking \keywords, a list of semantically replaceable keywords with prediction labels, by replacing their occurrences in issues with the specialized \textsf{[MASK]} token and (2) fine-tuning the parameters of the model by optimizing the prediction objective over the \labels of masks. Compared to \baselinecount state-of-the-art issue classifiers, varying from transformer architectures fine-tuned with different frameworks to autoregressive LLMs with different complexity and count of learnable parameters, the \classifier-trained classifier demonstrates a substantially better performance in detecting \secissue, outperforming the baselines by an average of 38.41\%, 48.10\%, and 48.21\% for precision, recall, and F1-score, over a corpus of 10,000 GitHub issues. Our best-performing classifier %, with an F1-score of 0.9880 at six training epochs, 
achieves 0.7123, 0.6860, and 0.6760 for precision, recall, and F1-score, when classifying 1,000 fully unseen issue reports. In this context, the precision and recall of the ready-made classifier over \textsf{[MASK]} token predictions are 0.7464 and 0.7020. Even in cases of zero occurrences of \keywords in unseen issues, our classifier can detect security-related submissions in real-time by incorporating predictions over the \textsf{[CLS]} head, as the fallback mechanism, with precision and recall equal to 0.6373 and 0.6341, respectively. The impressive success rate of \classifier makes it a practical training framework for producing robust and generalizable issue report classifiers for real-time triage of issue trackers.\\
\noindent \textbf{\underline{Future Work}:} We plan to improve the generalizability of \classifier by replacing its \textsf{[CLS]} prediction fallback mechanism with predictions over a new set of masked keywords that are compiled to only %capture the context in 
represent the issues that do not contain any occurrences of the initial list of \keywords. We also intend to evaluate the effectiveness of our training framework across different bidirectional transformers including variants of \bert. Finally, we plan to extend our framework to multi-label classification, not only detecting \secissue but also categorizing them based on their severity and the underlying vulnerabilities they hint at.

%% file: 09_data_availability.tex
\section{Data Availability}\label{sec:data-available}
To enable reproducibility, \classifier code and analysis scripts are available at \url{https://figshare.com/s/8fd18701f0de6d3429a7}.